\documentclass[conference]{IEEEtran}
\IEEEoverridecommandlockouts

\usepackage{cite}
\usepackage{amsmath,amssymb,amsfonts,amsthm}
\usepackage{algorithmic}
\usepackage{graphicx}
\usepackage{textcomp}
\usepackage{xcolor}
\def\BibTeX{{\rm B\kern-.05em{\sc i\kern-.025em b}\kern-.08em
    T\kern-.1667em\lower.7ex\hbox{E}\kern-.125emX}}

\ifCLASSINFOpdf
\else
\fi
\usepackage{soul}
\usepackage{subfigure}
\usepackage{tabularx,booktabs}
\usepackage{url}
\usepackage{multirow}
\usepackage{boldline}
\usepackage{pifont}
\theoremstyle{definition}
\newtheorem{definition}{Definition}

\begin{document}

\title{Discrete-Time Analysis of \\Wireless Blockchain Networks
\thanks{This work was funded by the IN CERCA grant from the Secretaria d'Universitats i Recerca del departament d'Empresa i Coneixement de la
Generalitat de Catalunya, and partially from the Spanish MINECO grant TEC2017-88373-R (5G-REFINE) and Generalitat de Catalunya grant 2017 SGR 1195.}
}


\author{\IEEEauthorblockN{Francesc Wilhelmi}
\IEEEauthorblockA{Centre Tecnol\`ogic de Telecomunicacions de Catalunya \\
(CTTC/CERCA)\\
Castelldefels, Barcelona, Spain \\
fwilhelmi@cttc.cat}
\and
\IEEEauthorblockN{Lorenza Giupponi}
\IEEEauthorblockA{Centre Tecnol\`ogic de Telecomunicacions de Catalunya \\
(CTTC/CERCA)\\
Castelldefels, Barcelona, Spain \\
lorenza.giupponi@cttc.es}
}


\maketitle

\begin{abstract}
Blockchain (BC) technology can revolutionize future networks by providing a distributed, secure, and unalterable way to boost collaboration among operators, users, and other stakeholders. Its implementations have traditionally been supported by wired communications, with performance indicators like the high latency introduced by the BC being one of the key technology drawbacks. However, when applied to wireless communications, the performance of BC remains unknown, especially if running over contention-based networks. In this paper, we evaluate the latency performance of BC technology when the supporting communication platform is wireless, specifically we focus on IEEE 802.11ax, for the use case of users' radio resource provisioning. For that purpose, we propose a discrete-time Markov model to capture the expected delay incurred by the BC. Unlike other models in the literature, we consider the effect that timers and forks have on the end-to-end latency.
\end{abstract}



\IEEEpeerreviewmaketitle

\section{Introduction}
One of the main challenges to face for Beyond 5G (B5G) networks is the need to foster the economic sustainability of the increasingly complex mobile networks. In this regard, a new economy of sharing has long been proposed by the main standardization bodies in contexts such as spectrum sharing, RAN/network sharing or network slicing. The exchange of services or resources among actors in these scenarios could be significantly automated and secured by the introduction of Blockchain (BC) to enable an economically driven and flexible network management. Through BC, providers can operate from the cloud to trade radio and network resources as-a-service, which enables novel trends such as autonomous network slicing or spectrum sharing with programmable and scientific trust. 

A BC is a decentralized, distributed record of transactions stored in a permanent and near inalterable way. Unlike traditional databases, typically administered by a central entity, BCs rely on a peer-to-peer (P2P) network of so-called miners that no single party can control. Authentication of transactions is achieved through cryptographic means and a mathematical consensus protocol that determines the rules by which the ledger is updated. This technology is useful to provide immutability, transparency, and security, which allows addressing multiple challenges in B5G networks. Besides the Bitcoin use case, BC technology has already been proposed to enhance multiple procedures in communications and networks~\cite{nguyen2020blockchain}, and some prominent examples are content delivery~\cite{herbaut2017model}, network slicing~\cite{backman2017blockchain}, RAN sharing~\cite{ling2019blockchain}, or network management~\cite{asheralieva2019distributed, maksymyuk2020blockchain}.  Nevertheless, BC entails several widely documented technological challenges, such as heavy computational costs, scalability issues, high latency, and energy consumption~\cite{BCscalability}. 

While most of the BC applications and deployments in the real world are designed considering a stable wired communication environment, in this paper, we focus on a more challenging wireless network as the communication infrastructure of our BC. Specifically, we propose a BC-enabled resource provisioning approach where end users can apply for radio and network resources on-demand, according to an as-a-service vision, without being bound in the long term to a specific operator. The needed exchange of transactions happens over a wireless network. Without loss of generality, we consider BC transactions to be exchanged over an IEEE 802.11ax network, to account also for the additional challenge of decentralization in the unlicensed spectrum. In such types of contention-based networks, BC users and miners need to compete for wireless resources to exchange transactions and blocks. 

In this context, we propose to model a wireless BC network through a batch-service queue based on a discrete-time Markov model that characterizes the latency introduced in the network. These aspects are key to the overall wireless system design. The longer it takes to distribute transactions and blocks in the BC network, the more the end-users performance is jeopardized, and the BC system becomes unstable, unreliable, and susceptible to forks. Based on the proposed model, we evaluate the resulting service latency using the Bianchi model~\cite{bianchi2000performance} for IEEE 802.11ax. In close spirit to this paper, we find the works in \cite{cao2019does,ngubo2020wi}, where the Bianchi model was also used to characterize IEEE~802.11 links through which the BC operation takes place. In contrast to the available literature, our model considers the effect of timers, whereby blocks are mined periodically, and forks, which can be seen as a side effect of the distributed consensus mechanism.

We demonstrate through simulation results that the proposed discrete-time Markov model perfectly captures the wireless BC behavior when also forks and timers are considered. Model analysis, as a function of different key parameters, allows gaining interesting insight on the dropping probability, the delay introduced by the BC (for which it is possible to identify an optimal block size), and the characteristic of the fork probability.
The rest of the paper is structured as follows: Section~\ref{section:related_work} overviews queue models for characterizing the delay incurred by the BC systems. Section~\ref{section:system_model} proposes a scenario where the BC is envisioned to enable resource provisioning in B5G. Section~\ref{section:queue_model} describes the proposed batch service queue model, and the corresponding results are provided in Section~\ref{section:performance_evaluation}. Finally, Section~\ref{section:conclusions} concludes the paper.

\section{Characterization of Blockchain:\\Related Work}
\label{section:related_work}

A BC is a type of distributed ledger technology (DLT) that compiles transactions in blocks that are sequentially chained one after the other. The record of the transactions is maintained across several computers which communicate through a P2P network. Nodes enabled to add transactions to the BC are referred to as miners. The consensus mechanism allows the BC to operate without the need to rely on a central trusted central authority. This automation of transactions is one of the most valuable features of BC for trading in future mobile networks.
BC was firstly introduced with the cryptocurrency Bitcoin~\cite{nakamoto2019bitcoin}. Lately, and enabled by the advent of smart contracts (i.e., computer programs that self-execute the terms of a contract when specific conditions are met) introduced by Ethereum~\cite{buterin2013ethereum}, the potential of BC was unleashed and incorporated into multiple domains \cite{cai2018decentralized}. 

The BC operation has been modeled in literature in multiple ways, ranging from analytical models to experimental tools~\cite{fan2020performance}. In the analytical domain, one of the tools gaining importance is batch service queuing~\cite{bar2007applications,chaudhry2012simple}, whereby packets leave the queue in batches, rather than individually. The batch property of these kinds of queues intuitively captures the BC idea of chains of blocks and has been already used to model the behavior of multiple technologies in communications \cite{bellalta2013performance}\cite{wen2016ruletris}\cite{kar2020throughput}.

Accordingly, the consensus-based confirmation procedure can be seen as a single-server batch service queue where packets (transactions) are served (mined) in batches (blocks). In this setting, the most relevant trade-off lies within the batch size and the fork probability. In both cases, the queuing delay is affected in a non-straightforward manner, and the most appropriate setting depends on factors like the arrivals rate, the number of miners, the implemented consensus mechanism, or the mining time (alternatively, the mining difficulty). Batch service queue models for BC have recently been introduced in~\cite{kawase2017transaction}, followed by the approaches in~\cite{kawase2018batch, li2018blockchain, geissler2019discrete, li2019markov, qi2020nash}. 

First, the work in~\cite{kawase2017transaction} proposed a batch service queue model to understand the stochastic behavior of the transaction-confirmation process in Bitcoin. Later, this model was extended in~\cite{kawase2018batch} to improve its accuracy for short block sizes, which was achieved by modeling transaction inter-arrivals with a general distribution. To do so, a matrix analytic method was used to derive the steady-state distribution of the queue model states, which were defined by the number of transactions in the queue just before new transaction arrivals. This model was validated with trace-driven simulations, and results showed that exponential-type distributions are accurate for estimating the mean confirmation time in Bitcoin. Similarly, the work in~\cite{li2018blockchain} attempted to overcome the complexity of the model in~\cite{kawase2017transaction, kawase2018batch} by separating the batch service queue into two separate processes, corresponding to block generation and block mining, respectively. This approach was later generalized by the same authors in~\cite{li2019markov}. Finally, a further evaluation was provided in~\cite{qi2020nash}, where a slightly different discrete-time queue model was used to study game-theoretical aspects related to the pricing associated with mining transactions.

Differently to the approaches discussed in~\cite{kawase2017transaction, kawase2018batch,li2018blockchain,li2019markov}, in this work we model queue states at departure instants, which was also considered in~\cite{geissler2019discrete}. Through this approach, the authors of~\cite{geissler2019discrete} were able to provide a more detailed performance evaluation of BC, on Ethereum. Unlike what proposed until the date, we further extend the batch service queue to incorporate the key effects of timers and forks within the BC operation. To the best of our knowledge, none of these features has been considered before in any BC model. We believe that these features are fundamental to any BC model for multiple reasons. First, timers are important to break the dependence of the mining procedure on the arrivals rate, which allows guaranteeing a maximum waiting delay, even if blocks are not filled with transactions (i.e., blocks are mined periodically). Disregarding the effects of timers leads to poor accuracy when deriving the queue status or the expected delay~\cite{claeys2013tail}. Second, the forks are a dramatic source of instability in BC. The appropriate modeling of their behavior ensures high accuracy and fidelity to capture the BC network's real dynamics. 

\section{System Model}
\label{section:system_model}
\subsection{Overview of the scenario}
End users are not necessarily bound to a specific operator in the long term but, at any moment, can trigger requests to get specific data services. Based on the BC technology, the User Equipment (UE) devices issue smart contracts that facilitate automatic, transparent, and auditable mechanisms whereby operators and service providers fulfill UE requests. The smart contracts (indicating service indicators such as duration, maximum latency, or throughput) are registered in a public BC, which enables flexible and decentralized on-demand network access. 


In the proposed scenario, UE smart contracts are submitted to the closest AP~\cite{ngubo2020wi}, which, besides providing radio access to the UEs, act as peer nodes in the BC P2P network. An AP is responsible to carry out tasks such as broadcasting transactions, storing a copy of the BC, achieving consensus, or mining blocks. In this paper, we focus on analyzing the delays to get the UE requests to reach the mobile operators, leaving for future work the study of how the service is provided by the selected operator to the UEs.

\subsection{Blockchain delays}
In the proposed BC-enabled communication environment, peers (miners) generate candidate blocks with transactions (service requests) from the UEs, which are mined once the block size $S_B$ is reached, or when a timer $T_\text{w}$ expires. To mine a block, peer nodes run a given consensus mechanism such as Proof-of-Work (PoW) and then propagate it over the P2P network. Upon successful block propagation, the winning miner is allowed to append the candidate block to the BC. 

To assess the feasibility of the BC operation, it is important to identify its associated delays, especially when running over wireless links. Despite providing security and decentralization capabilities, the delay associated to the BC may impact the performance of the underlying applications if the service latency is too high (e.g., V2X communications), and can also introduce instability in the BC, due to the increased risk of forks. Considering the proposed BC-enabled procedure, we identify the following steps for a UE request to be served: 
\begin{enumerate}
    \item \textbf{Smart contract upload:} first, the UE selects the closest peer node to submit a smart contract to indicate the requirements of the requested service. Then, P2P nodes verify and propagate the received transactions over the P2P network. At this stage, it is important to collect valid, non-conflicting transactions (e.g., to prevent double-spending). The delay for uploading smart contracts ($T_\text{up}$) depends on the capabilities of the communication links between UEs and peer nodes, which may vary depending on the technology (e.g.,~IEEE~802.11) and the environmental conditions (e.g., distance, interference, frequency of operation).
    \item \textbf{Queuing:} Valid transactions are queued before being included into candidate blocks. The queuing delay ($T_\text{q}$) depends on the number of arrivals, the mining rate, and the block size. Transactions are included into the candidate block till the block size is reached, or the timer $T_\text{w}$ expires. The queuing delay is characterized in Section~\ref{section:queue_model} by modeling the BC as a batch-service queuing system. 
    \item \textbf{Block generation:} Peer nodes execute the consensus algorithm until finding the block's nonce, or until receiving a new valid mined block. As widely considered in the literature~\cite{decker2013information, biais2019blockchain}, the block generation time ($T_\text{bg}$) is modeled as an exponential random variable with parameter $\mu$, which is assumed to be the same for all the miners.\footnote{In Bitcoin's PoW, the mining time is fixed to an average of ten minutes by adjusting difficulty based on miners' computational power.}
    \item \textbf{Block propagation:} The winning miner generates a new block, based on its candidate block, which is distributed through the P2P network with a delay of $T_\text{bp}$. Once the mined blocks are propagated, the included transactions are put into effect.
\end{enumerate}

Notice that steps (2)-(4) may be repeated as a result of forks. A fork can be seen as a side effect of the consensus mechanism and the communication and distribution delay required for the synchronization of the BC, which leads to splitting the BC into different paths, each one representing a potentially different status of the BC. If a miner different than the winner (i.e., the miner that mined the block in the shortest amount of time) succeeds in generating a new block while the first successful block is still being propagated, some uninformed miners may mistakenly add the second block to their version of the BC. Assuming that $M$ miners with the same exponentially distributed block generation rate $\mu$ start the mining process synchronously, the fork probability can be defined as:\footnote{The time between mining events is modeled as a Poisson inter-arrival process following an exponential distribution.}
\begin{equation}
\begin{split}
    p_\text{fork} &= 1 - \prod_{\forall i \neq w} \Pr(T_\text{bg}(i) - T_\text{bg}(w) > T_\text{bp}) \\&= 1 - e^{-\mu (M-1)T_\text{bp}},
\end{split}
\label{eq:fork_probability}
\end{equation}
where $T_\text{bg}(w)$ is the winner's block generation delay. It can be noticed that, for a fixed mining capacity, the higher the propagation delay, the higher the fork probability and the more unstable the BC will be.

To sum up, and taking into account the effect of forks, the BC operation incurs the following end-to-end delay to make a UE request effective:
\begin{equation}
    T_\text{BC} = T_\text{up} + \frac{(T_\text{q} + T_\text{bg} + T_\text{bp})}{1-p_\text{fork}}
\end{equation}

\section{Batch Service Queue Model}
\label{section:queue_model}

\begin{figure}[ht!]
\centering
\includegraphics[width=.9\linewidth]{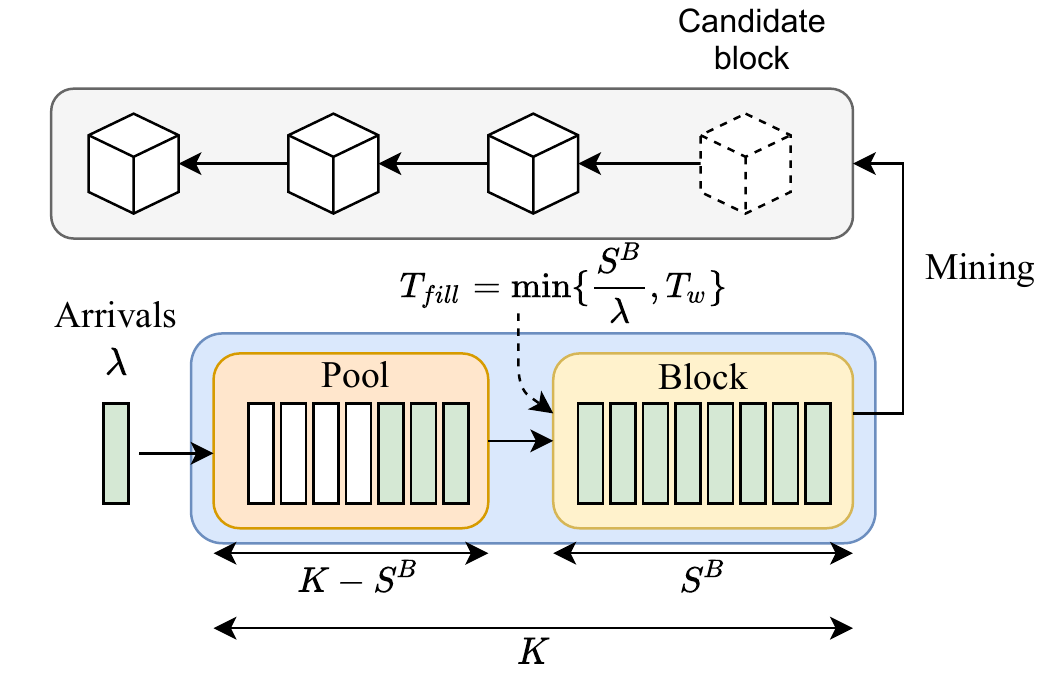}
\caption{Blockchain procedure through batch-service queuing.}
\label{fig:batch_service_queue}
\end{figure}

We consider a batch-service queuing approach to model the time a transaction spends in the ledger before being validated. In a finite-length $M/M^s/1/K$ queue, $s$ denotes the number of packets served altogether. The batch characteristic is useful to represent the BC operation, where the mining procedure starts when the block size $S_B$ is reached, or when the timer ($T_\text{w}$) expires (see Fig.~\ref{fig:batch_service_queue}). 

In particular, we are interested in calculating the average number of packets in the queue ($\text{E}[Q]$), which allows deriving the expected queuing delay ($\text{E}[D]$), including both $T_\text{q}$ and $T_\text{bg}$. For that purpose, we first model the queue status at departure instants $\boldsymbol{\pi}^d$, as done in~\cite{bellalta2013performance}. Then, we derive the queue's steady-state distribution $\boldsymbol{\pi}^s$ by solving a finite number of system of equations. We use Little's law~\cite{shortle2018fundamentals} to obtain the expected delay, so that $\text{E}[D] = \frac{\text{E}[Q]}{\lambda(1-p_b)}$, where $\lambda$ is the total arrival rate, considering all the UEs, and $p_b = \pi_K^s$ is the blocking probability (the probability of discarding a transaction because the queue of length $K$ is full). Finally, the expected queue occupancy is obtained as $\text{E}[Q] = \sum_{k=0}^K k \pi_k^s$. 

\subsection{Departures Distribution}

\begin{figure}[ht!]
\centering
\includegraphics[width=.9\columnwidth]{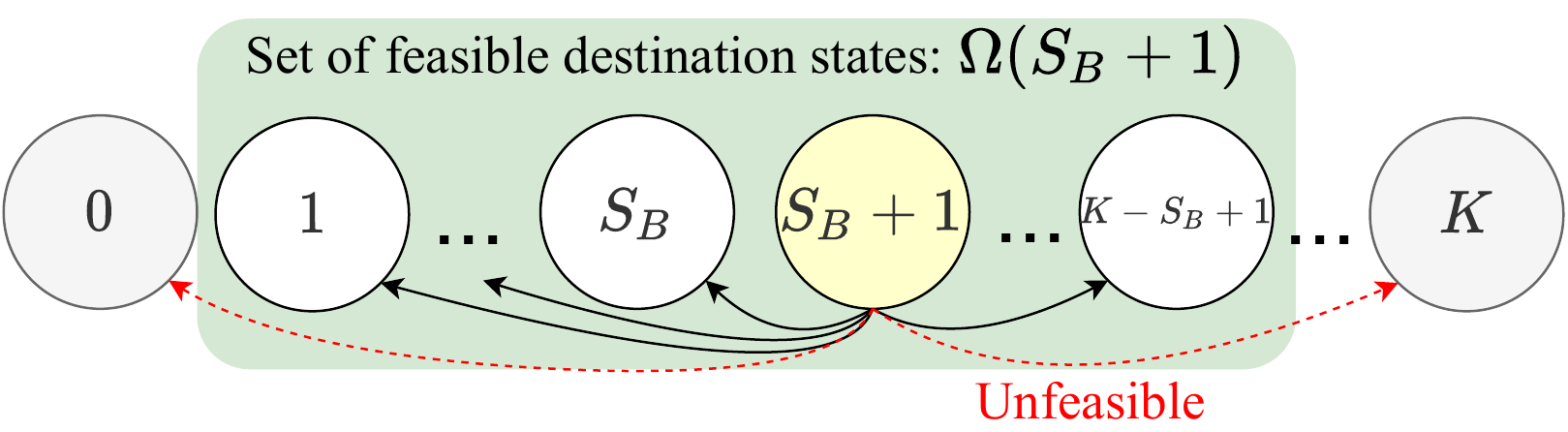}
\caption{Discrete-time Markov chain representing the feasible departure states from reference state $S_B + 1$.}
\label{fig:markov_chain}
\end{figure}

We define the status of the batch service queue at the $n$-th departure instant as:
\begin{equation}
  q_n = 
    \min \Big\{ q_{n-1} + a(q_{n-1}) - s(q_{n-1}), K - s(q_{n-1}) \Big\},
\end{equation}
where $q_{n-1}$ is the status of the queue after the previous departure, $a(q_{n-1})$ is the number of new arrivals during the last inter-departure epoch, $s(q_{n-1})$ is the number of packets served in a batch, and $K$ is the queue length. Based on this, we derive the departure probabilities by considering the set of feasible states $\boldsymbol{\Omega}(q_n)$ that are reachable from any state $q_n$. To further illustrate this, Fig.~\ref{fig:markov_chain} shows an example of the possible transitions from state $q_n = S_B+1$. From that state, $a(q_{n-1})$ is limited by $K$ (the maximum reachable state is $K-S_B+1$), and the minimum reachable state is upper bounded by $q_n - s(i)$.

In general, the set of feasible destination states $\boldsymbol{\Omega}(i)$ that can be reached from state $i$ is $j \in \big[ [i-s(i)]^+, K-s(i) \big]$. Notice that any transition to $j > K-s(i)$ is not possible because the transitions to be served are kept in the queue until the block is mined. Under the assumption that arrivals and departures follow independent Poisson and exponential distributions, respectively, and by being aware of the possible inter-departure transitions, the probability $p_{i,j}$ for reaching a state $j$ from $i$ is as follows:

\begin{equation}
  \resizebox{\linewidth}{!}{%
  $p_{i,j} = \begin{cases}
    \int_{0}^\infty e^{-\lambda t} \frac{(\lambda t)^{j}}{j!} \cdot \mu e^{-\mu t} dt, & [i - s(i)]^+ \leq j < K - s(i)
    \\
    \\
    1-\sum^{K-s(i)-1}_{l=0} p_{i,l}, & j = K - s(i)
    \\
    \\
    0, & \text{otherwise}
    \end{cases}$
    }
\end{equation}

The first case ($[i - s(i)]^+ \leq j < K - s(i)$) indicates that the maximum queue size is not reached with new transactions. In turn, the second case ($j = K - s(i)$) refers to the situation in which the queue is filled because the system received, at least, $K - s(i)$ transactions during the inter-departure epoch. Other transitions are unfeasible. Since arrivals and departures follow independent Poisson and exponential distributions, respectively, the probability $p_{i,j}$ for $[i - s(i)]^+ \leq j < K - s(i)$ can be rewritten as:
\begin{equation}
    p_{i,j} = \frac{\mu}{\mu + \lambda} \Big(\frac{\lambda}{\mu + \lambda}\Big)^{j-(i-s(i))}
    \label{eq:pij}
\end{equation}

The departure probability distribution $\boldsymbol{\pi}^d$ can be obtained from the transition-probability matrix $\boldsymbol{\text{P}}$, which is built with probabilities $p_{i,j}$. In particular, $\boldsymbol{\pi}^d$ can be calculated by solving $\boldsymbol{\pi}^d = \boldsymbol{\pi}^d \boldsymbol{\text{P}}$, provided that $\boldsymbol{\pi}^d \boldsymbol{1}^T = 1$. 

\subsection{Steady-state Distribution}
From the departures distribution $\boldsymbol{\pi}^d$, the steady-state queue occupancy distribution $\boldsymbol{\pi}^s$ can be derived thanks to the Poisson Arrivals See Time Averages (PASTA) property~\cite{gross1998fundamentals} (or random observer property). The PASTA property states that, for Poisson processes, the queue status at any arbitrary time, $\boldsymbol{\pi}^s_k$, is equal to the status probability at which random arrivals find the queue, $\boldsymbol{\pi}^a_k$. 

To compute the steady-state distribution, we must first consider the time the queue spends in each state $i$, $T(i)$, which, as shown earlier, is divided into filling and mining periods. In particular, the time $T(i)$ the system spends until the next departure from state $i$ is given by:
\begin{equation}
    T(i) = \min \{ T_\text{w}, [S_B - i]^+/\lambda \big)\} + \frac{1}{\mu},
\end{equation}
which depends on the probability $\tau(i)$ that the timer $T_\text{w}$ expires before the minimum block size $S^B$ is reached from state~$i$. In particular, for any state $i<S^B$, the timer expiration probability is given by: 
\begin{equation} 
    \tau(i) = \sum_{j=i}^{S^B-1}  e^{-\lambda T_\text{w}} \frac{(\lambda T_\text{w})^j}{j!}
\end{equation}

With the departures distribution $\boldsymbol{\pi}^d$ and expected time spent in a given departure state, the steady-state probability of a state $k \in [0,K]$ is derived as in Eq.~\ref{eq:pi_s}. Notice that, to include the effect of the timer for $k < K$, we need to consider the exact amount of packets that can be observed during the filling procedure, which conditions the probability of observing state $k$ in the mining phase. In particular, from any departure state $i$, we consider the probability of observing exactly $n$ packets (denoted by P($n|\tau$)) independently, given that $n < S^B - i$ and that the timer expires.

\begin{figure*}[!t]
\begin{equation} \label{eq:pi_s}
\pi_k^s = \begin{cases}
    \frac{1}{\lambda \text{E}[T]} \sum_{i=0}^k \pi_i^d \bigg( \overline{\tau}(i) \Big( \sum_{j=k-s(i)+1}^{K-s(i)} p_{i,j} \Big) + \tau(i) \Big( \sum_{j=i}^{S^B-1} \text{P}(n=j-i|\tau) \big(\sum_{l=k-s(j-i)+1}^{K-s(j-i)} p_{j,l}\big) \Big) \bigg), & k < K
    \\
    \\
    1-\sum^{K-1}_{i=0} \pi_{i}^s, & k = K
    \end{cases}
\end{equation}
\hrulefill
\end{figure*}

\subsection{Considering forks}
To capture the effect of forks with the model, we need to consider that forks affect the mining procedure in two main ways:
\begin{enumerate}
    \item The fact that several miners work concurrently affects the mining rate.
    \item Transactions involved in a fork are re-added to the pool of unconfirmed transactions (i.e., the queue). Throughout this work, we assume the worst-case scenario where all the transactions involved in a fork are re-added.
\end{enumerate}

Regarding implication 1), we need to consider the first order statistics of the individual exponential random variables characterizing the mining time.
\begin{definition}
Let $X_i$ be an exponential random variable and $X_{(k)}$ be the \textit{k}-th smallest instance among $n$ occurrences $X_1, ..., X_n$. Assuming that $X_{(i)}$ is a reordered sequence of $X_i$, s.t. $X_{(1)} \leq X_{(2)} \leq ... \leq X_{(n)}$, then $X_{(1)}$ is a random variable $\text{Exp}(n\lambda)$ whose distribution is $f_{X_{(1)}}(x) = n\lambda e^{-n \lambda x}$, and whose expected value is $\text{E}[X_{(1)}] = \frac{1}{n\lambda}$. 
\end{definition}

With this, we capture the competition among miners (the first one to mine a block, wins), which leads to characterizing the mining time as a joint exponential distribution $\text{Exp}(n\lambda)$.

Now, assuming synchronization among miners and that the same transactions are included in any block replica, we address implication 2) by modifying the behavior of the queue when a departure occurs. This affects the transition probability matrix $\text{P}$ and the blocking probability $p_b$, since transactions may not leave the queue after a departure epoch, provided that forks occur. In particular, the number of served transactions from departure state $i$, namely $s(i)$, is defined as follows:
\begin{equation}
    s(i) = \bar{p}_\text{fork}\cdot\min\{i,S^B\} + p_\text{fork}\cdot t_{b,w} \not\in \mathcal{T}_{b,i\neq w},
\end{equation}
where $t_{b,w} \not\in \mathcal{T}_{i\neq w}$ is the set of transactions included in winner's block $b$ that have not been affected by the fork (i.e., the set of non-overlapping transactions). Assuming that miners include exactly the same set of transactions to every block replica, $t_{b,w} = \emptyset$.

\section{Performance Evaluation}
\label{section:performance_evaluation}
In this Section, we validate the proposed batch-service queue model and evaluate the performance of BC IEEE 802.11ax networks in cellular-based random deployments.


\subsection{Simulation Parameters}
To characterize the IEEE 802.11ax links used by the BC nodes, we use the Bianchi's Distributed Coordination Function (DCF) model \cite{bianchi2000performance}, which allows estimating the saturation throughput $\Gamma$ of DCF overlapping devices as a two-step probability function:
\begin{equation}
    \Gamma = \frac{p_{\text{slot,s}}\text{E}[L]}{\text{E}[T_\text{slot}]},
\end{equation}
where $p_{\text{slot,s}}$ is the probability of transmitting successfully at a given slot, $\text{E}[L]$ is the average payload length, and the average duration of a generic slot $\text{E}[T_\text{slot}]$. Considering that Request To Send / Clear To Send (RTS/CTS) is enabled, the duration of each type of slot (empty, successful, and collision slots, respectively) is as follows:
\begin{equation}
  \resizebox{\columnwidth}{!}{%
   $\begin{cases}
       T_\text{slot,e} = 9 \mu s\\
       T_\text{slot,s} = T_{RTS} + 3\cdot T_{SIFS} + T_{CTS} + T_{DATA} + T_{ACK}\\
       T_\text{slot,c} = T_{RTS} + T_{DIFS}
   \end{cases} $
   }
\end{equation}

Table \ref{tbl:simulation_parameters} details the parameters used in this Section.	
\begin{table}[ht!]
\centering
\caption{Simulation parameters}
\label{tbl:simulation_parameters}
\resizebox{\columnwidth}{!}{%
\begin{tabular}{|r|l|l|l|}
\hline
 & \multicolumn{1}{c|}{\textbf{Parameter}} & \multicolumn{1}{c|}{\textbf{Description}} & \multicolumn{1}{c|}{\textbf{Value}} \\ \hline
\multirow{2}{*}{\textbf{Depl.}} & $N$ & Num. of APs/cells & 19 \\ \cline{2-4} 
 & $R$ & Cell radius & 10 m \\ \hline
\multirow{11}{*}{\textbf{PHY}} & $B$ & Bandwidth & 20 MHz \\ \cline{2-4} 
 & $F_c$ & Carrier frequency & 5 GHz \\ \cline{2-4} 
 & $MCS$ & Modulation and coding scheme & 0-11 \\ \cline{2-4} 
 & $S$ & Single-user spatial streams & 1 \\ \cline{2-4} 
 & $T_{PH}$ & PHY header duration & 20 $\mu$s \\ \cline{2-4} 
 & $T_{OFDM}$ & OFDM symbol duration & 4 $\mu$s \\ \cline{2-4} 
 & $P_t$ & Transmit power & 20 dBm \\ \cline{2-4} 
 & $PL_0$ & Loss at the reference dist. & 5 dB \\ \cline{2-4} 
 & $\alpha$ & Path-loss exponent & 4.4 \\ \cline{2-4} 
 & $\sigma$ & Shadowing factor & 9.5 \\ \cline{2-4} 
 & $\gamma$ & Obstacles factor & 30 \\ \hline
\multirow{9}{*}{\textbf{MAC}} & $CW_{min/max}$ & Min/max contention window & 32 \\ \cline{2-4} 
 & $L_{D/ACK}$ & Data and ACK lengths & 12.000 / 32 bits \\ \cline{2-4} 
 & $L_{RTS/CTS}$ & RTS and CTS lengths & 160 / 112  bits \\ \cline{2-4} 
 & $L_{MH}$ & MAC Header & 320 bits \\ \cline{2-4} 
 & $N_a$ & Max. A-MPDU & 1 frame \\ \cline{2-4} 
 & $T_{PPDU}$ & Max. PPDU duration & 5,4884 $\mu$s \\ \cline{2-4} 
 & $T_{DIFS/SIFS}$ & DIFS/SIFS duration & 34 / 16 $\mu$s \\ \cline{2-4} 
 & $T_e$ & Empty slot duration & 9 $\mu$s \\ \cline{2-4} 
 & $CCA$ & CCA threshold & -82 dBm \\ \hline
\multirow{3}{*}{\textbf{BC}} & $\mu$ & Mining capacity & 15 blocks/s \\ \cline{2-4} 
 & $L_T$ & Transaction length & 3.000 bits \\ \cline{2-4} 
 & Q & Batch service queue length & 10 packets \\ \hline
\end{tabular}%
}
\end{table}

As for signal propagation effects, we have considered a standard log-distance path loss model with shadowing effects. In particular, the loss observed between nodes $i$ and $j$ is:
\begin{equation}
    PL(d_{i,j}) = PL_{0} + 10\alpha \log_{10}(d_{i,j}) + \frac{\sigma}{2} + \frac{\gamma}{2} \frac{d_{i,j}}{10}
\end{equation}

\subsection{Numerical Results}
\begin{figure}[ht!]
\centering
\subfigure[$T_w = 0.5$s, forks disabled]{\includegraphics[width=0.49\columnwidth]{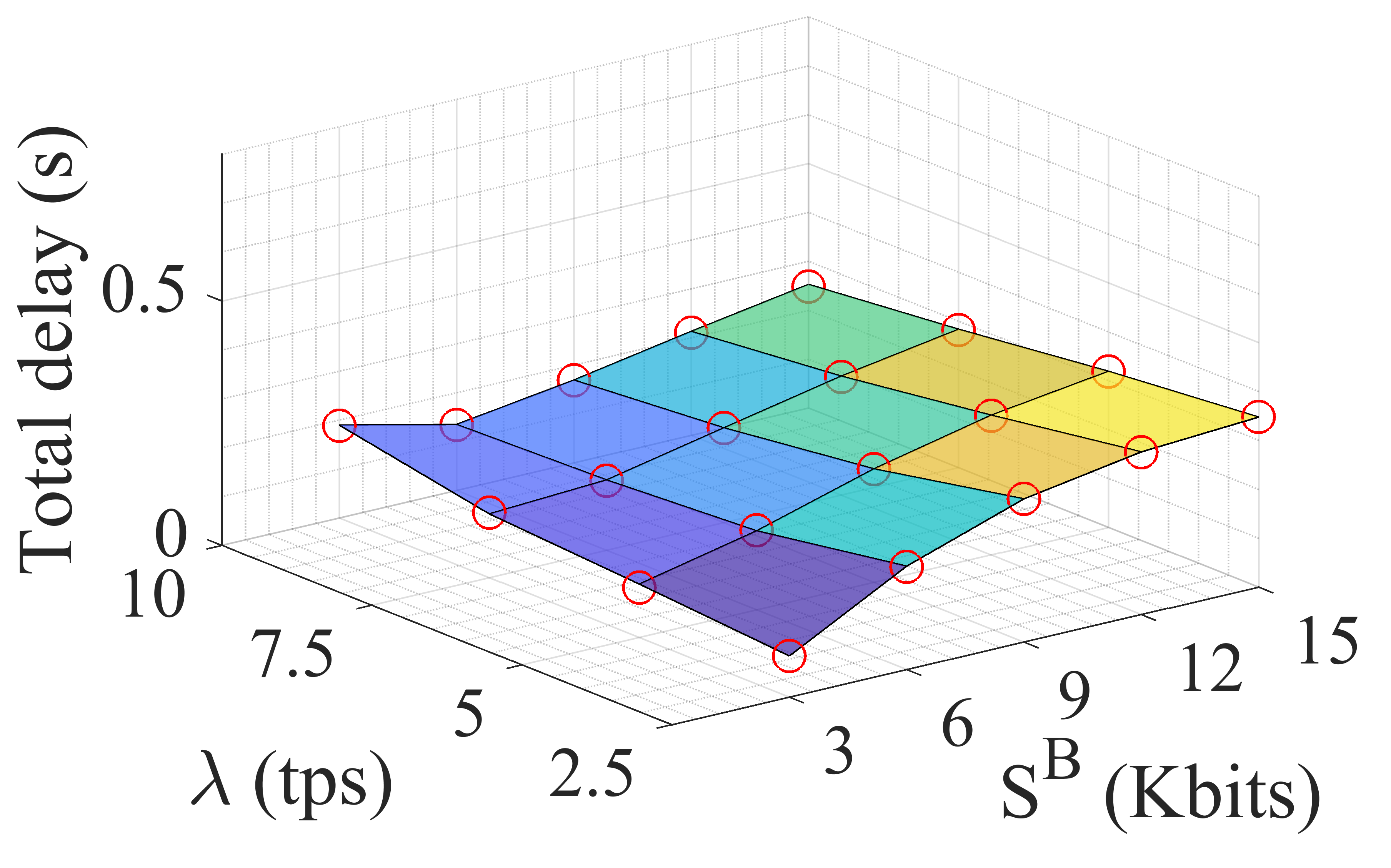}} 
\subfigure[$T_w = 2$s, forks disabled]{\includegraphics[width=0.49\columnwidth]{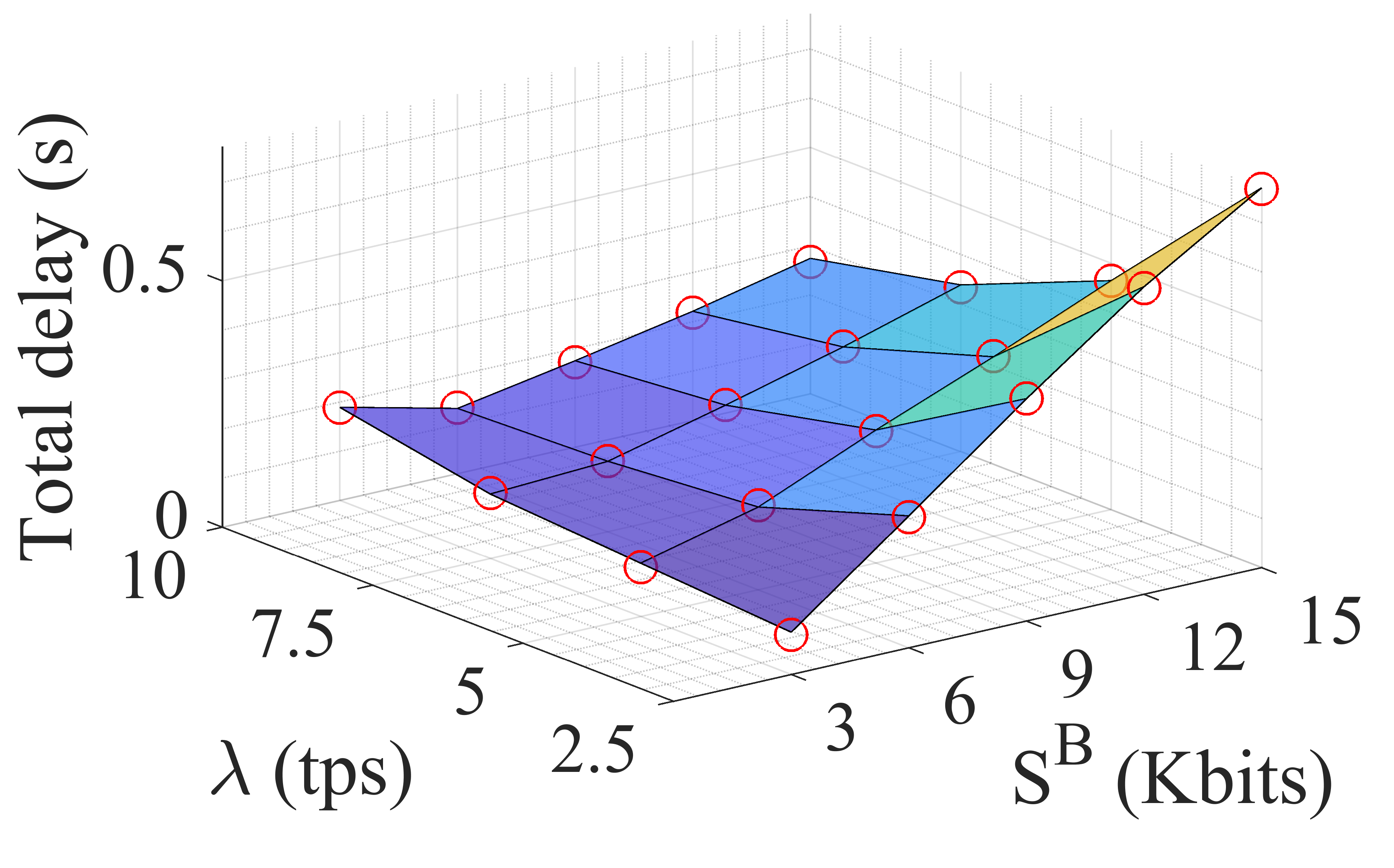}} 
\subfigure[$T_w = 0.5$s, forks enabled]{\includegraphics[width=0.49\columnwidth]{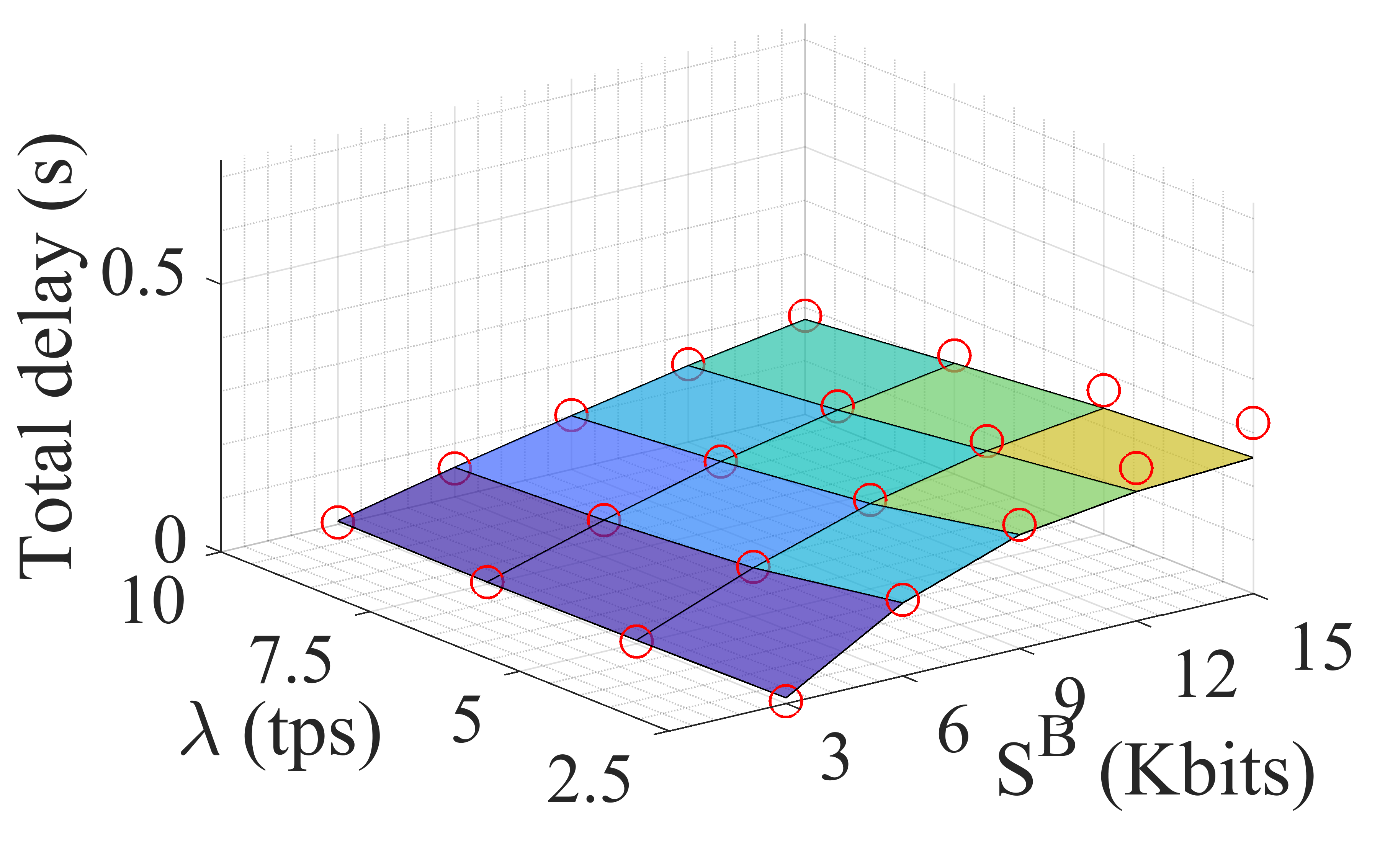}} \subfigure[$T_w = 2$s, forks enabled]{\includegraphics[width=0.49\columnwidth]{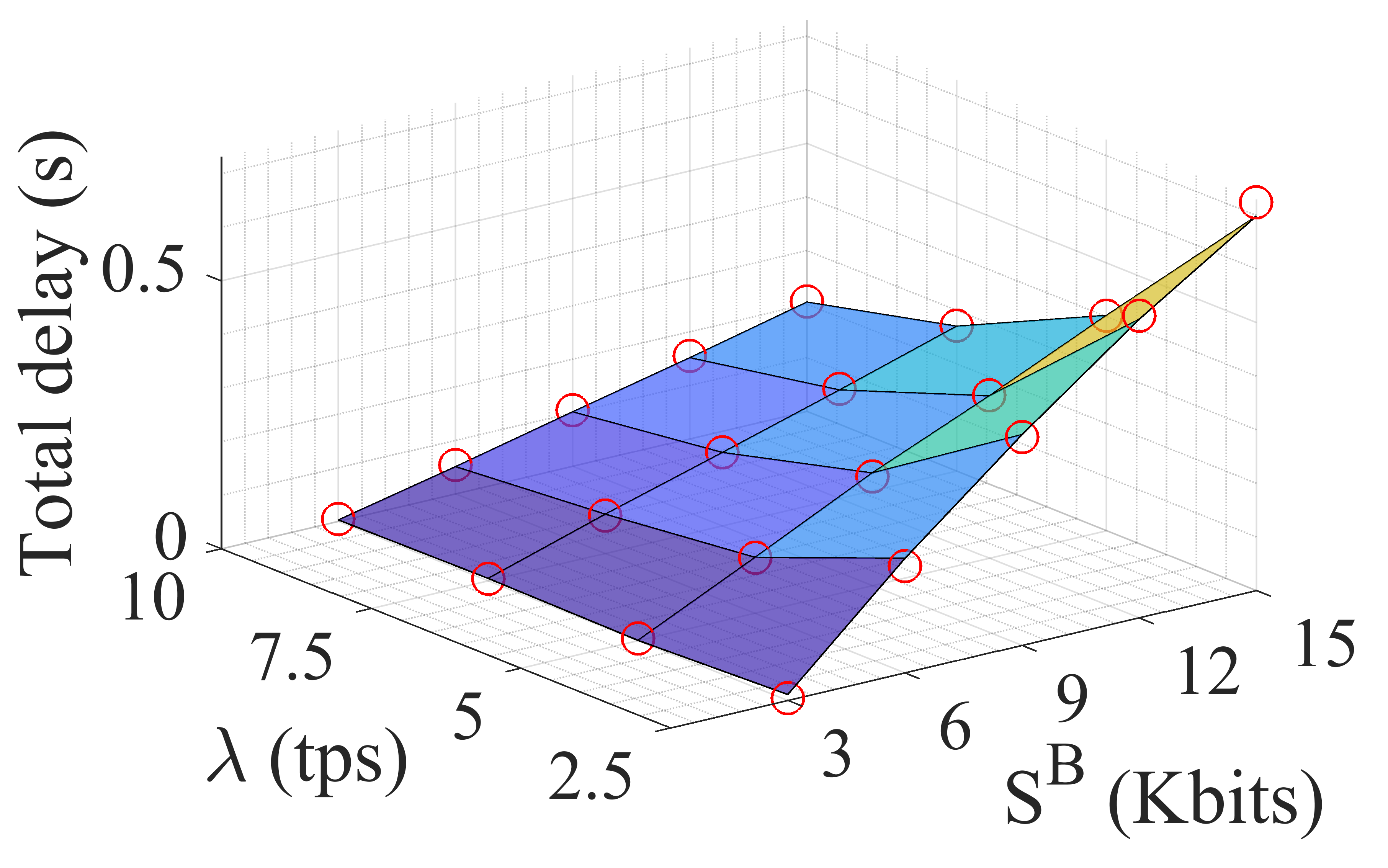}}
\caption{Queuing delay for different UE arrivals rates ($\lambda$) in transactions per second (tps) and block size ($S^B$) in Kbits. The presence or not of forks is also evaluated. The solid areas correspond to model results, while the red circles correspond to simulation results.}
\label{fig:total_delay}
\end{figure}

We start validating our queue model with simulation results.\footnote{The batch-service queue simulator used for validations is open-access~\cite{batchsim}.} Fig.~\ref{fig:total_delay} shows both model and simulation results of the queuing delay, for different user arrivals rates and block sizes. For all $\lambda$ and $S^B$ values, two different timers values ($T_w=0.5$s and $T_w=2$s) are selected to showcase the situations in which the timer is likely or not to expire, respectively, before filling a block. Besides, the presence or not of forks is included in the validation. As Fig.~\ref{fig:total_delay} shows, model results match with simulations, thus demonstrating that the model properly captures the effect of both timers and forks.

\begin{figure}[ht!]
\centering
\subfigure[Queuing delay]{\includegraphics[width=\columnwidth]{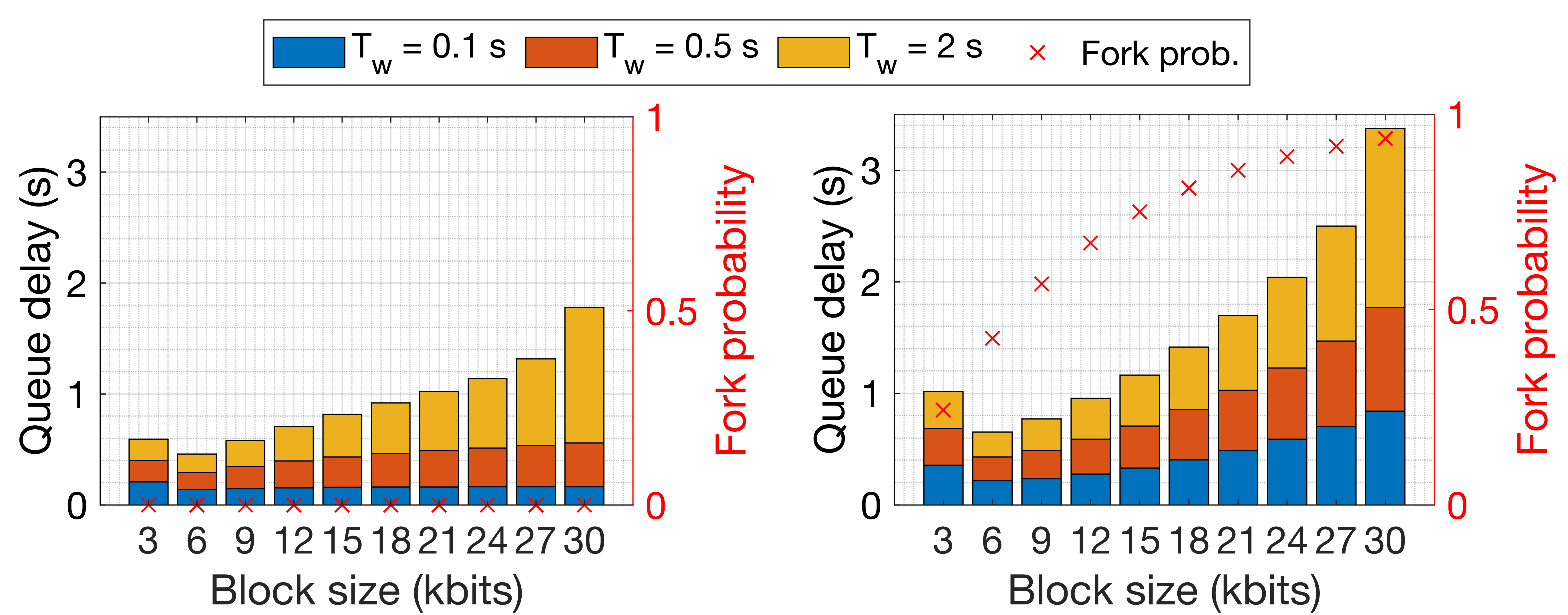}\label{fig:delay_analysis1}} 
\subfigure[Drop probability]{\includegraphics[width=\columnwidth]{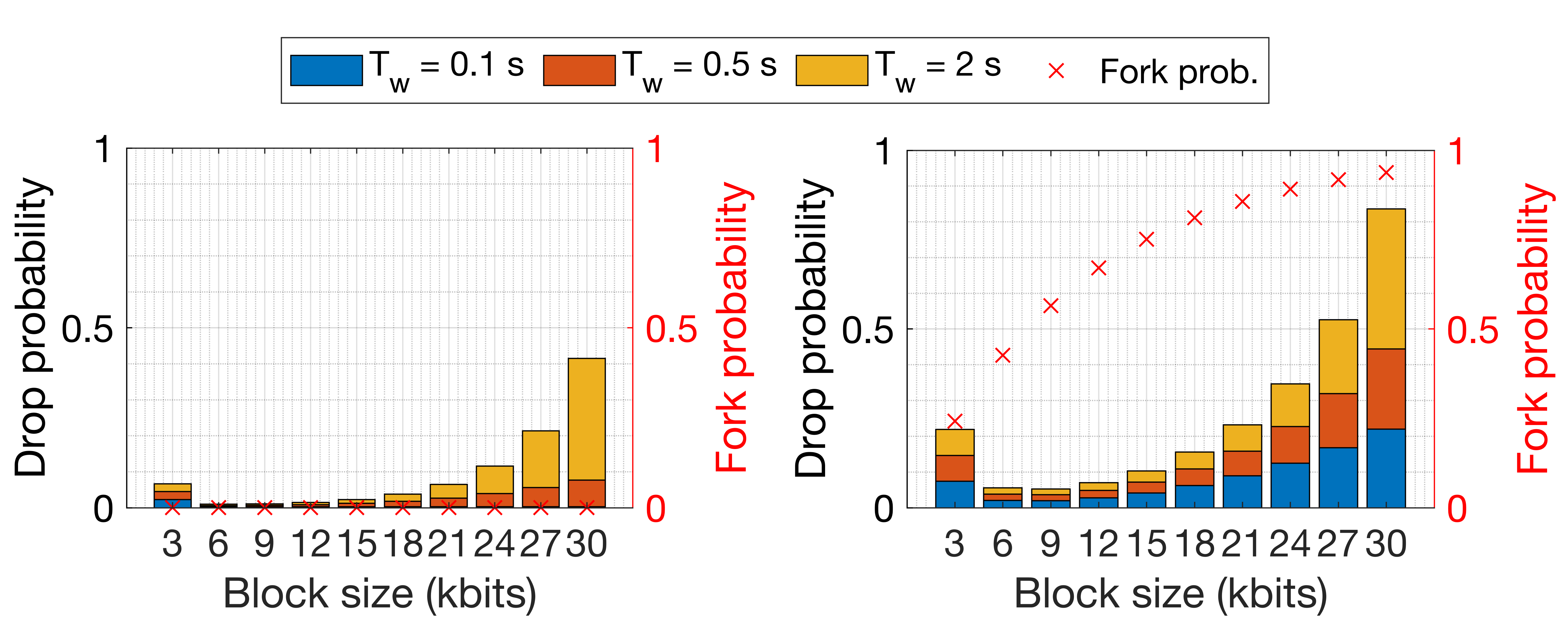}\label{fig:delay_analysis2}} 
\caption{Analysis of the queuing delay and drop probability for different block size and timer values.}
\label{fig:delay_analysis}
\end{figure}

To better illustrate the impact of the different parameters ($\lambda$, $S^B$ and $T_w$), Fig. \ref{fig:delay_analysis1} and Fig.~\ref{fig:delay_analysis2} show the queuing delay and the drop probability, respectively, observed with and without forks. For each considered block size $S^B$, we have averaged the results for different user arrivals, namely $\lambda=\{2.5, 5, 7.5, 10, 12.5, 15\}$ arrivals per second. This time, only model results are provided.

As shown, the queue delay is optimum at $S^B = 6$~Kbits. From that point, the delay increases with the block size, even if forks are enabled or not. As for the timer values, we observe that the queue delay is significantly reduced when the timer is low (e.g., $T_w = 0.1$ s). The timer is particularly useful to reduce BC latency when the number of arrivals is low and blocks are not filled with transactions at the necessary speed. Nevertheless, setting the timer to a low value contributes to generating more overhead and may also be counterproductive in terms of delay (notice that the mining time is insensitive to the number of transactions in a block).

As for the fork probability, it increases with the block size until saturation. Specifically, the higher the block size, the higher the block propagation time, which, for a fixed mining rate, contributes to making forks more likely to occur. An increase in the fork probability has a great impact on the number of drops, thus packet losses are noticed even for small timer values (i.e., even if the queue is not full).

\begin{figure}[ht!]
\centering
\subfigure[Forks disabled]{\includegraphics[width=\columnwidth]{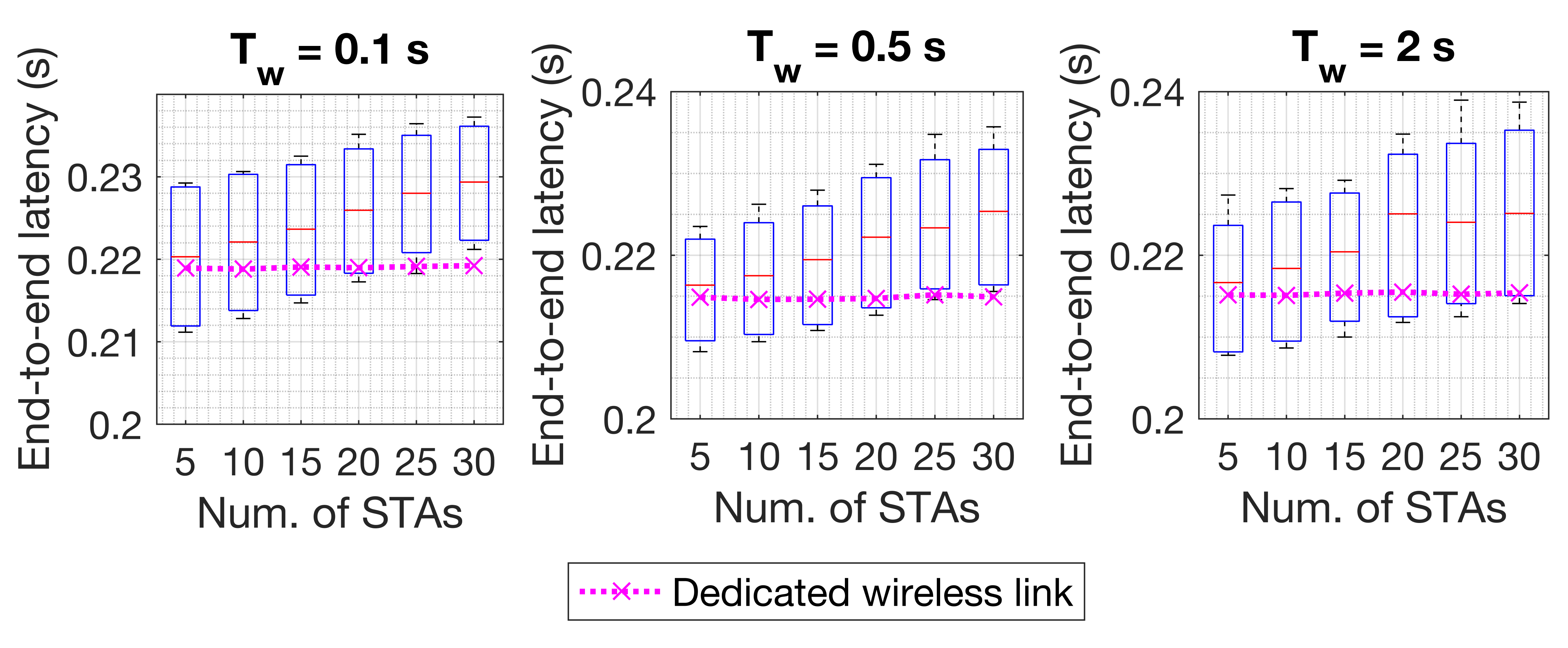}\label{fig:density1}} 
\subfigure[Forks enabled]{\includegraphics[width=\columnwidth]{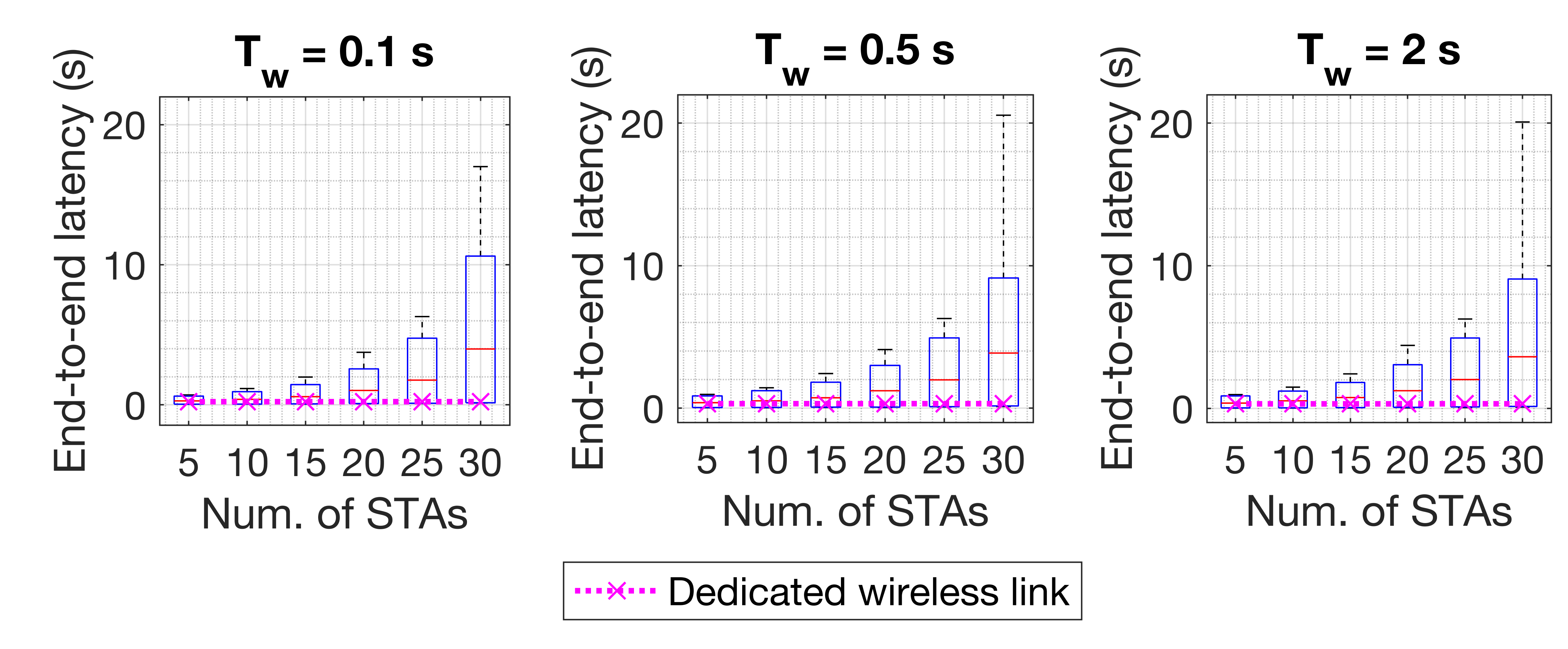}\label{fig:density2}} 
\caption{Analysis of the end-to-end latency for different network densities. Results are shown for shared (boxplots) and dedicated (purple dashed lines) IEEE 802.11ax links.}
\label{fig:density_analysis}
\end{figure}

Finally, we assess the performance of the BC-enabled resource provisioning application for different user densities. Fig.~\ref{fig:density_analysis} shows the end-to-end latency of the wireless BC for validating transactions. We have considered up to 30 concurrent STAs with full-buffer traffic. The block size is set to the fixed value of 6 Kbits so that the queue delay is minimized, whereas the transactions arrival rate ($\lambda$) is set to 7.5 arrivals per second. Moreover, for each considered density, we have simulated 10 random deployments for averaging purposes. The delay obtained over shared IEEE 802.11ax links is compared to the one where the P2P network operates over dedicated IEEE 802.11ax wireless channels.

As shown in Fig.~\ref{fig:density1}, for the ideal case where no forks occur, the use of shared IEEE 802.11ax links is affordable even if network density increases. This is due to the low amount of network resources used by the BC application. In contrast, when forks are considered (see Fig.~\ref{fig:density2}), the latency increases dramatically with the network density. The block propagation delay increases when the number of concurrent users is high. This leads to a high fork probability, which negatively impacts the number of attempts for validating transactions. 

\section{Conclusions}
\label{section:conclusions}
In this paper, we have envisioned a future wireless system where users are not bound to a specific operator and apply for resources through smart contracts. BC allows secure and decentralized transactions among users and operators, but its latency performance represents a challenge for the service perception of the UE and for the stability of the BC, which will increasingly suffer from forks as the latency augments. While BC systems are normally based on wired P2P networks, in this paper we foresee the BC to be based on a wireless infrastructure. We have focused on IEEE 802.11ax, to add the additional challenge of contention in the performance evaluation. With the aim to investigate the wireless BC performance, 
we have analytically modeled the BC through a batch queue serving system based on a discrete-time Markov model. Results have revealed that our model perfectly captures the behaviors of the wireless BC, including timers and forks, which is novel in literature, and shown interesting optimal points of the block size and behaviors of the fork probability, as a function of both the network and BC are designed. 


\ifCLASSOPTIONcaptionsoff
\newpage
\fi

\bibliographystyle{IEEEtran}
\bibliography{bibliography}

\end{document}